\documentstyle[prl,aps,epsfig]{revtex}
\newlength{\figwidth}
\title{ Percolation model for  nodal domains of chaotic  wave functions}
\author{ E. Bogomolny and C. Schmit}
\address{Laboratoire de Physique Th\'eorique et Mod\`eles Statistiques \dag \\
 Universit\'e de Paris-Sud, B\^at. 100, 91405 Orsay Cedex, France}
\begin{document}

\twocolumn

\maketitle
\begin{abstract}
Nodal domains are regions where a function has definite sign. In
\cite{Uzy} it is conjectured that the distribution of nodal domains
for quantum eigenfunctions of chaotic systems is universal. We propose a
percolation-like model for description of these nodal domains which permits
to calculate all interesting quantities analytically, agrees well with
numerical simulations, and due to the relation to
percolation theory opens the way of deeper understanding of the structure 
of chaotic wave functions.
\end{abstract}

\vspace{.5cm}

05.45.Mt, 05.45.Df, 61.43.Hv, 64.60.Ak

\vspace{.5cm}


In a recent paper \cite{Uzy} Smilansky et al. consider the
following problem. Let $\Psi(x,y)$ be a real eigenfunctions of a
2-dimensional quantum problem. The equation $\Psi(x,y)=0$
determines a set of  nodal lines which separate nodal domains 
where $\Psi(x,y)$ is of opposite signs. In \cite{Uzy} it is
argued that the distribution of the number of these regions for high
excited states is (i) universal for integrable as well as for chaotic
models but (ii) it clearly distinguishes between these two types of models.

For  chaotic (billiard) systems it is conjectured in \cite{Uzy} that this
distribution   coincides with the distribution of nodal domains for 
Gaussian random functions which are known to give a good description of wave
functions of chaotic systems \cite{Berry}:
\begin{equation}
\Psi(x,y)=\sum_{m=-\infty}^{\infty} C_m\Psi^{(0)}_m(x,y),
\label{2}
\end{equation}
where $\Psi^{(0)}_m(x,y)=J_{|m|}(kr)e^{ i m\phi}$
form  the standard basis for billiard problems, $k$ is the momentum, $E=k^2$,
and $C_m=C_{-m}^{*}$ are  independent random variables with  Gaussian
distribution. Only numerical calculations of this distribution have been 
performed in \cite{Uzy}.

The purpose of this letter is to demonstrate that nodal domains  of random 
functions (\ref{2}) (and, consequently,  wave functions of generic chaotic
systems \cite{Berry}, \cite{Uzy}) can be described by a simple 
percolation-like model where all interesting quantities can be calculated 
analytically.  The model  permits also to apply ideas and methods
developed within the percolation theory to the field of quantum chaos.

\begin{figure}[ht]
\begin{center}
\epsfig{file=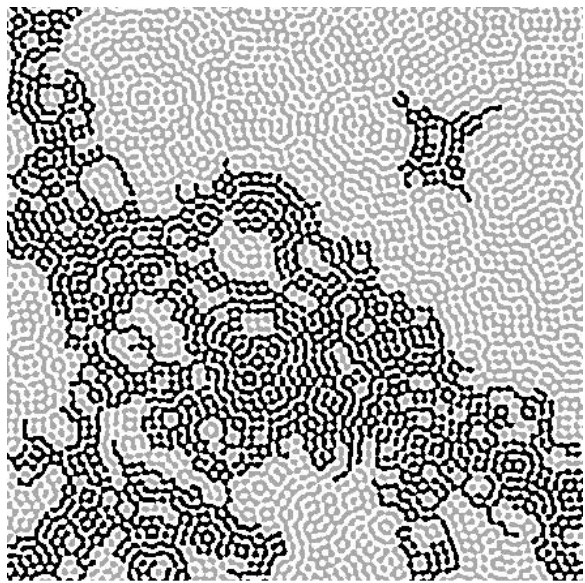,  width=\figwidth}
\end{center}
\caption{Nodal domains for a realization of random function (\ref{2}) with
  $k=100$. Two different connected domains are highlighted.}
\label{fig9}
\end{figure}

To understand how the nodal domains look like we give
in Fig.~\ref{fig9} their picture for random function (\ref{2})
with $k=100$. The figure corresponds to a square window of the size $L=4$ which 
contains 907 connected nodal domains. The largest of them and the
largest of the domains which do not touch the boundary are highlighted. 

Our  first step is to calculate the mean number of zeros of random functions
(\ref{2}) along a straight line (say the vertical one). This can be achieved
by noting that, if the size in $y$-direction is $L_y$,  the approximate
quantization condition reads $\bar{k}_yL_y\approx \pi m$
where $m$ is an integer and $\bar{k}_y$ is the mean square momentum along the
$y$-axis, $\bar{k}_y^2=k^2/2$. Therefore when $x$ is fixed
\begin{equation}
\bar{\rho}(y)=\frac{m}{L_y}=\frac{k}{\pi \sqrt{2}}.
\label{6}
\end{equation}

The rigorous derivation of this relation  using the method
of\cite{Bogomolny} will be given elsewhere \cite{Schmit}. 
The same answer, of course, can be obtained for the mean density of nodal
lines along any other straight lines \cite{footnote1}.

These simple arguments show that the mean density of intersections of any
straight line with  nodal lines of random functions (\ref{2}) is the same as
in (\ref{6}) and  nodal lines of these random functions can in the 
{\bf mean} be considered as forming an approximate rectangular lattice whose
total number of sites is asymptotically
\begin{equation}
N_{tot}\approx \frac{k^2 A}{2\pi^2}=\frac{2}{\pi} \bar{N}(E),
\label{7}
\end{equation}
where $A$ is the area of the billiard and $\bar{N}=A E/4\pi$
is the mean number of levels (for billiards) with energy less than $E$.

\begin{figure}[ht]
\begin{center}
\epsfig{file=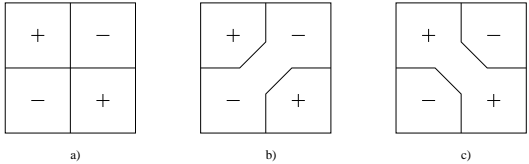, angle=270,  width=\figwidth}
\end{center}
\vspace{-1.5cm}
\caption{a) True nodal crossing. b) and  c) Avoided nodal crossings.}
\label{fig2}
\end{figure}

But this simple picture can be valid only in the mean. When
$\Psi(x,y)=\bar{\Psi}(x,y)+\delta \Psi(x,y)$
where function $\bar{\Psi}(x,y)$ has a  crossing of nodal lines as in
Fig.~\ref{fig2}a, the addition of a small correction $\delta \Psi(x,y)$ will
change, in general, the true crossing to one of two possible avoided
crossings as in Figs.~\ref{fig2}b and \ref{fig2}c.

\vspace{-2cm}

\begin{figure}[ht]
\begin{center}
\epsfig{file=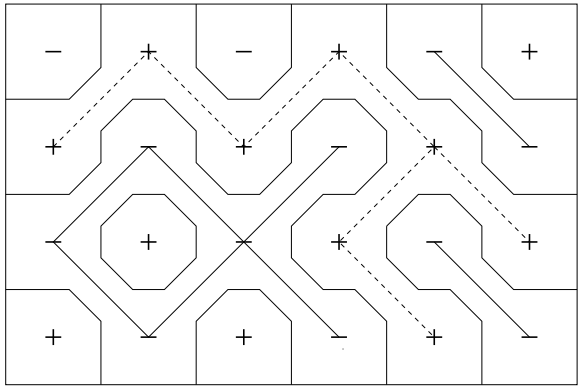, width=\figwidth}
\end{center}
\caption{A realization of random percolation-like process. Plus and minus 
form two dual lattices. Solid and dashed lines indicate  graphs for 
respectively negative and positive dual lattices.}
\label{fig3}
\end{figure}

Consequently, one can  conjecture that the distribution of nodal domains for 
random functions is the same as for the following random percolation-like  
process. 
Let us consider a rectangular lattice with the total number of sites
$N_{tot}=2 \bar{N}(E)/\pi$ as in (\ref{7}). Each line crossing with probability
$1/2$ is changed either to the avoided crossing as in Fig.~\ref{fig2}b 
or to the one as in Fig.~\ref{fig2}c. 
These rules give a well defined random percolation-like process. 
One realization of such a process is presented in Fig.~\ref{fig3}.

The original lattice gives rise to two dual lattices called below positive
and negative whose vertices are in the centers of  regions where
our function is positive or negative (see Fig.~\ref{fig3}) and whose size,
$a$, coincides with de Broglie wave length (cf. (\ref{6})): $a=2\pi/k$.
Any realization of the above mentioned random process uniquely defines
two graphs on these lattices, (which we call also  positive and 
negative) with the following properties (i) their vertices coincide with 
the vertices of the corresponding lattice, (ii) 
their edges join together the connected components  of this  lattice. 
(A point is also a component of the  graph.)

One can choose arbitrarily a  graph on one lattice
(say negative) and any of such graphs will correspond to an
allowed realization and vice versa. Therefore our random  process is
determines mostly by the bond percolation model on one of dual lattices (see e.g.
\cite{percolation}) where with probability $1/2$ one connects 2 nearby sites 
by a bond.  

The number of  connected  nodal domains coincides 
with the sum of the numbers of different components of both positive 
and negative  graphs. As in \cite{Uzy} we  first are interested in the 
distribution of these numbers. To compute this quantity 
(unusual for the percolation)  it is convenient to connect this model 
with the Potts model (see e.g. \cite{Wu}) similarly as it was done in
\cite{Duplantier} for a slightly different problem.

Let $n_{\pm}$ be the numbers of connected  components of positive and
negative graphs. The generating function of their sum is
\begin{equation}
Z(x)=\sum_{realizations}x^{n_{-}+n_{+}},
\label{zxy}
\end{equation}
where variable $x$ plays the role of the fugacity.

The negative and positive graphs, by construction, are dual to each
other \cite{footnote2}
and  their properties are interrelated. In particular (see e.g. \cite{Wu}
p. 242) $n_{+}=C_{-}+1$
where $C_{-}$ is the number of independent circuits on the negative (dual to
the positive) graph. According to the Euler relation this quantity
can be expressed as follows $C_{-}=b_{-}+n_{-}-N_{-}$
where  $b_{-}$ is the number of bonds, $n_{-}$ is the number of connected
components and $N_{-}$ is the number of vertices of the negative graph. 

These relations permit to express the generation function (\ref{zxy})
through the properties of only negative graph, $G_{-}$,  
\begin{equation}
Z(x)=x^{1-N_s}\sum_{G_{-}}x^{b_{-}}(x^2)^{n_{-}}
\label{z}
\end{equation}
where we take into account that $N_{-}$ equals the total number of sites of
negative lattice, $N_s=N_{tot}/2$.

But this quantity is directly connected with the partition sum of the 
Potts model \cite{Wu}, \cite{Baxter}. The later can be defined for an 
arbitrary  graph by the formal sum
\begin{equation}
Z_{Potts}(v,q)=\sum_{G}v^{b(G)}q^{n(G)},
\label{potts}
\end{equation}
where the summation is performed over all  graphs, $G$,  which cover the
original graph. $b(G)$ is the number of bonds of this graph, $n(G)$
is its number of connected components. $q$ is the number of states of the
Potts model, $v=e^K-1$ is a parameter related with the inverse temperature $K$.

Comparing (\ref{z}) and (\ref{potts}) one gets
\begin{equation}
Z(x)=x^{1-N_s}Z_{Potts}(x,x^2).  
\label{zmain}
\end{equation}
The last sum corresponds to the Potts model in the critical point $v^2=q$;
for large rectangular lattice and $q<4$ it was computed analytically  
\cite{Baxter} 
\begin{equation}
\lim_{N_s\rightarrow \infty}\frac{1}{N_s}\ln Z_{Potts}(x,x^2)= \log x+ f(x),
\end{equation}
where
\begin{equation}
f(x)=\int_{-\infty}^{\infty}\frac{dt}{t}\tanh \mu t \frac{\sinh (\pi
  -\mu)t}{\sinh \pi t},
\label{f(z)}
\end{equation}
and the parameter $\mu$ $(0<\mu <\pi/2)$ is related to the fugacity $x$ as
follows: $\cos \mu =x/2$.

The expansion of this integral into the series on $z$ gives the number of
nodal domains with a fixed number of components, $N_n$. With exponential
accuracy
\begin{equation}
  \sum_{n=1}^{\infty} N_nz^n=z \exp (N_s f(z)).
\end{equation}
We are interested in the behavior of $N_n$ for large $n$ near the maximum of $N_n$ 
with fixed number of sites. One has
\begin{equation}
N_n=\frac{1}{2\pi i} \oint \frac{dz}{z^{n}}\exp N_s f(z),
\end{equation}
where the integration is performed over a contour around zero.
Assuming that
$n$ is large and using the saddle point method one obtains
$N_n\propto \exp \Phi (n,z_c)$,
where $\Phi (n,z)=N_s f(z) -n\ln z$
and the saddle point $z_c$ is determined from the equation
$\partial \Phi(n,z)/\partial z|_{z=z_c} =0$.
The maximum of $N_n$  corresponds to $n=\bar{n}$ for which $\Phi (n,z_c)$ is
maximal. Simple calculation  shows that it appears when $z_c=1$  and 
\begin{equation}
\frac {\bar{n}}{N_s}=z\frac{d f(z)}{dz}|_{z=1}.
\end{equation}
Expanding $\Phi (n,z_c)$ near $n=\bar{n}$ up to the second order
and computing all necessary integrals (the details will be given elsewhere
\cite{Schmit}) one finds that the total number
nodal domains in the lattice with $N_{tot}=2\bar{N}(E)/\pi$ sites (cf.
(\ref{7})) near the maximum has  Gaussian distribution where the mean
number of nodal domains, $\bar{n}$, and their variance, $\sigma^2$, are
related to the mean number of levels in the following way  
\begin{eqnarray}
&&\bar{n}=\bar{N}(E)\frac{3\sqrt{3}-5}{\pi},
\label{nmean}\\
&&\sigma^2= \bar{N}(E)(\frac{18}{\pi^2}+\frac{4\sqrt{3}}{\pi}-\frac{25}{2\pi}).
\label{sigmamean}
\end{eqnarray}
These relations are the main analytical result of this note. They 
demonstrate that the
distribution of nodal domains for random functions is universal (Gaussian)
and depends only on the mean number of levels exactly  as it is conjectured in
\cite{Uzy} with explicitly calculable constants.

\begin{figure}[ht]
\begin{center}
\epsfig{file=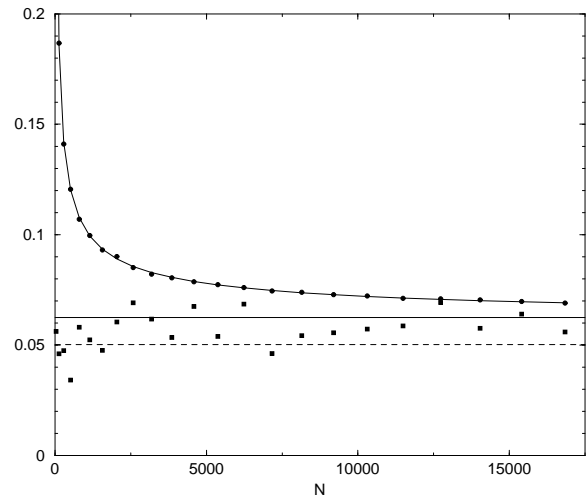, angle=270, width=\figwidth }
\end{center}
\caption{Mean values of the nodal domains (dots) and their variances (squares)
  for random functions divided by $N$ versus $N$. The solid and 
  dashed horizontal  lines represent  theoretical asymptotic values 
  (\ref{nmean}) and (\ref{sigmamean}) respectively. }
\label{fig4}
\end{figure}

In Fig.~\ref{fig4} we present the comparison between the numerical
calculation of the mean value and the variance for the random functions
(\ref{2}) for different values of $k$ plotted as functions of $N=\bar{N}(E)$.
The dots and squares represent the ratios $\bar{n}(N)/N$  and $\sigma^2(N)/N$ 
averaged over  100 realizations. Within statistical errors  numerical results
agree with the  predictions (\ref{nmean}) and (\ref{sigmamean}).
Small errors may also be related
to finite size effect  as the infinite sum in (\ref{2}) has been cut at 
$|m|\leq 2 kL$ where $L=4$ is the side of the chosen  square window.  

The close relation of our model  with the bond percolation model at the
critical point $p=1/2$ permits to apply all results of percolation theory 
(see e.g. \cite{percolation} and references therein)
for description of nodal domains of random functions.

In particular the percolation theory predicts that the distribution of the
areas, $s$,  of clusters (= connected nodal domains), $n(s)$, should have 
power behavior
\begin{equation}
n(s) \propto s^{-\tau}
\label{tau}
\end{equation}
where the Fisher exponent $\tau=187/91$ \cite{percolation} p.52. 

\begin{figure}[ht]
\begin{center}
\epsfig{file=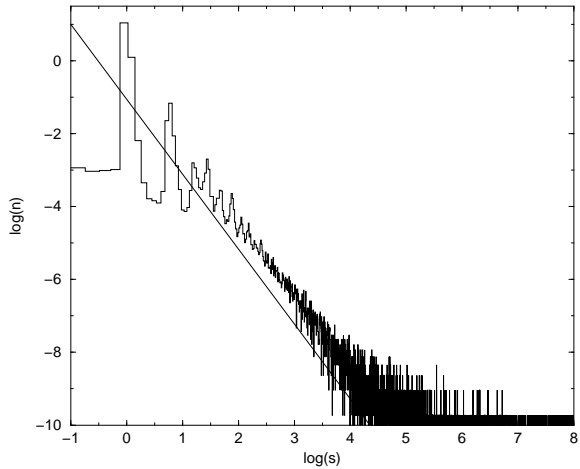, angle=270, width=\figwidth}
\end{center}
\caption{Distribution of nodal domain areas.
  The solid line has the slope $\tau=187/91$ predicted by the  percolation
  theory.}
\label{fig5}
\end{figure}

In Fig~\ref{fig5} we present the results of numerical calculations of
this quantity for random functions (\ref{2}) with $k=115$ which are in 
a rather good agreement with the percolation theory prediction.
In this figure the $y$-axis represents the  number of nodal
domains divided by $\bar{N}(E)$ and the area along the
$x$-axis is measured in the unit of $s_{min}=\pi (j_1/k)^2$ where $j_1$ is
the first zero of the Bessel function, $J_0(j_1)=0$, which according to the
Rayleigh inequality \cite{Polya} is  the smallest possible (with fixed $k$) 
area. After such scaling the results for random
functions with different $k$ are practically superimposed. The existence of
discrete set of smallest possible areas leads to  pronounced oscillations at
small $s$ in Fig~\ref{fig5}.

Another interesting quantity is the fractal dimension of the nodal domains.
In our percolation model it  coincides with the fractal dimension of
critical percolation clusters which is known to be equal to $D=91/48$ 
\cite{percolation} p.52.
To find numerically fractal dimension of a domain
it is convenient to put it on  a grid of  squares of side $R$ and count the
number of crossing of the region with the grid. When $a\ll R\ll l$ where $l$
is the size of the domain and $a$ is the size of the mean lattice:
$a=2\pi/k$, one expects 
\begin{equation}
n\propto R^{-D}
\label{D}
\end{equation}
and the exponent $D$ is the fractal dimension.

\begin{figure}[ht]
\begin{center}
\epsfig{file=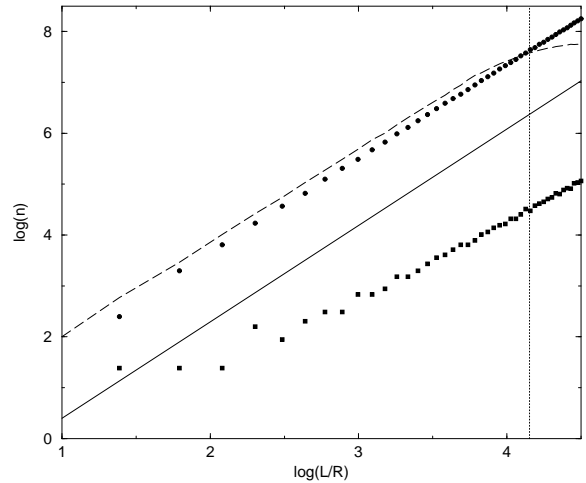, angle=270, width=\figwidth}
\end{center}
\caption{Number of intersections of nodal domains with square
  grid of size $R$. Dotted vertical line indicates the  mean
  lattice size. Circles and squares correspond
  respectively to large and small highlighted domain in Fig.~\ref{fig9}.
  The dashed line: results of numerical calculations for the largest cluster 
  in the percolation-like model. 
  The solid line: percolation theory prediction with the exponent $D=91/48$.}
\label{fig6}
\end{figure}

In Fig.~\ref{fig6} we present numerical verification of this relation 
for the two  nodal domains with $k=100$ highlighted in Fig.~\ref{fig9} and for
the largest cluster in the proposed percolation-like model with number of sites 
given by (\ref{7}). It is clearly seen that both domains have close fractal 
dimensions  which agree well with  simulations in our percolation-like model  and 
the percolation theory prediction.

To summarize,  we developed a simple percolation-like model to describe 
the nodal domains for random functions. Its main advantage is that all relevant
quantities can be computed analytically. By using the relations with the
Potts model we demonstrated that nodal domains are distributed according to 
Gaussian distribution whose mean value and  variance are proportional to 
the mean staircase function with explicitly calculated parameters
(\ref{nmean}) and (\ref{sigmamean}). Our results clearly indicate that the distribution of nodal
domains for random functions is in the same universality class as critical
bond percolation which permits us to predict different critical  exponents 
like the Fisher exponent for the distribution of the nodal domain areas 
(\ref{tau}) and its fractal dimension (\ref{D}). 

Many different generalization of the model considered are possible. We
mention only the possibility to use non-critical percolation model for
description of level domains  of random functions,
$\Psi(x,y)=\epsilon$, with $\epsilon \neq 0$.

The authors are greatly indebted to U. Smilansky for discussing the paper
\cite{Uzy} prior the publication. It is a pleasure to thank  O. Bohigas, 
J. Jacobsen, X. Campi, S. Nechaev, and B. Duplantier for fruitful discussions.


\begin{thebibliography}{99}
\bibitem{footnote1} Using the same method  higher correlation
  functions of the density of zeros along different lines can be computed
  analytically as well.
\bibitem{footnote2} For finite graphs the dual graph has to include  one
  vertex at infinite distance (see e.g. \cite{Wu}). As it is only a boundary
  effect we shall ignore it.
\end{thebibliography}
\end{document}